\documentclass[conference]{IEEEtran}
\IEEEoverridecommandlockouts

\usepackage{cite}
\usepackage{amsmath,amssymb,amsfonts}
\usepackage{algorithmic}
\usepackage{graphicx}
\usepackage{textcomp}
\usepackage{xcolor}
\usepackage{multirow, tabularx, makecell}
\usepackage{booktabs}
\usepackage[font=normalsize]{caption}
\def\BibTeX{{\rm B\kern-.05em{\sc i\kern-.025em b}\kern-.08em
    T\kern-.1667em\lower.7ex\hbox{E}\kern-.125emX}}

\usepackage[hyphens]{url}
\usepackage{hyperref}

\begin{document}

\title{SV-Mixer: Replacing the Transformer Encoder with Lightweight MLPs for Self-Supervised Model Compression in Speaker Verification\\
\thanks{This work was supported by the National Research Foundation of Korea (NRF) grant funded by the Korea government. (MSIT) (2023R1A2C1005744)}
}

\author{
\IEEEauthorblockN{
Jungwoo Heo, 
Hyun-seo Shin, 
Chan-yeong Lim, 
Kyo-won Koo, \\
Seung-bin Kim, 
Jisoo Son, 
Ha-Jin Yu$^\dagger$\thanks{$^\dagger$Corresponding author}
}
\IEEEauthorblockA{
University of Seoul, Korea \\
\texttt{jungwoo4021@gmail.com, gustjtls123@naver.com, cksdud585@naver.com, kkwr0504@uos.ac.kr}\\ 
\texttt{kimho1wq@naver.com, hojasu07@naver.com, hjyu@uos.ac.kr}
}
}

\maketitle

\begin{abstract}
Self-supervised learning (SSL) has pushed speaker verification accuracy close to state-of-the-art levels, but the Transformer backbones used in most SSL encoders hinder on-device and real-time deployment. 
Prior compression work trims layer depth or width yet still inherits the quadratic cost of self-attention. 
We propose SV-Mixer, the first fully MLP-based student encoder for SSL distillation. 
SV-Mixer replaces Transformer with three lightweight modules: Multi-Scale Mixing for multi-resolution temporal features, Local-Global Mixing for frame-to-utterance context, and Group Channel Mixing for spectral subspaces. 
Distilled from WavLM, SV-Mixer outperforms a Transformer student by 14.6\% while cutting parameters and GMACs by over half, and at 75\% compression, it closely matches the teacher's performance. 
Our results show that attention-free SSL students can deliver teacher-level accuracy with hardware-friendly footprints, opening the door to robust on-device speaker verification. 
\end{abstract}


\section{Introduction}
Recent advances in self-supervised learning (SSL) have driven substantial progress in speech representation learning, with models such as HuBERT~\cite{hubert}, Wav2Vec 2.0~\cite{wav2vec}, and WavLM~\cite{wavlm} widely adopted across various speech tasks, including automatic speech recognition~\cite{asr1, asr2} and speaker verification~\cite{sv1, sv2, sv3, sv4}. 
Despite a three-year gap in development, SSL-based models from \textbf{2022}~\cite{wavlm} remain on par with the latest supervised systems introduced in \textbf{2025}~\cite{redimnet}, underscoring the lasting strength of SSL in speaker verification. 
Nevertheless, SSL-based speaker verification remains relatively underexplored compared to the extensive progress made in supervised approaches. 
This disparity can be attributed to two key challenges. 
First, training large-scale SSL models demands considerable computational resources, which makes fine-tuning and adaptation difficult in resource-constrained environments. 
Second, even once training is complete, most SSL models remain too large and computationally intensive for deployment on mobile or embedded platforms, limiting their use in real-time or resource-constrained environments. 

To mitigate these limitations, recent studies have explored various strategies for compressing SSL models. 
While notable progress has been made through techniques such as layer-wise distillation~\cite{distilhubert, fithubert}, hint-based supervision~\cite{lighthubert,star}, and pruning~\cite{parp}, most approaches still preserve the Transformer architecture in the student model. 
Consequently, \textbf{previous work remains tied to the Transformer architecture and its inherent limitations}, particularly the quadratic complexity of self-attention that incurs computational cost~\cite{tay2022efficient, att1, att2}. 

The structural limitations of Transformer architectures have been widely recognized across domains, motivating active exploration of alternative designs. 
In computer vision, MLP-based models such as MLP-Mixer~\cite{mlpmixer} and CycleMLP~\cite{cyclemlp} have demonstrated competitive performance in computer vision, while offering simpler designs and improved scalability. 
Similar trends are emerging in speech applications as well. 
For instance, MLP-SVNet~\cite{mlpsvnet} adopts a purely MLP-based architecture for speaker verification in a supervised setting, while ConvMixer-based models~\cite{convmixer} have been successfully applied to keyword spotting under noisy and low-resource conditions. 
These studies collectively highlight the feasibility of attention-free architectures across modalities. 

Recent work on compressing self-supervised learning (SSL) models for speaker verification largely remains tied to Transformer backbones, incurring the quadratic cost of self-attention at deployment. 
At the same time, MLP-based vision models have shown that attention-free designs can retain strong representational power with far lower complexity. 
Building on these observations, we introduce \textbf{SV-Mixer}, a fully Transformer-free student encoder for lightweight SSL-based speaker verification. 
Directly inserting off-the-shelf MLP architectures such as MLP-Mixer or CycleMLP into a distillation pipeline, however, proved inadequate in our preliminary tests, motivating a design tailored to the requirements of distillation. 
SV-Mixer therefore replaces self-attention with three lightweight yet expressive mixing modules. 
\textbf{Multi-Scale Mixing (MSM)} processes features at multiple temporal resolutions to capture both short- and long-range speech dynamics. 
\textbf{Local-Global Mixing (LGM)} couples frame-level and utterance-level representations, enriching contextual information. 
\textbf{Group Channel Mixing (GCM)} partitions channels into groups, increasing modeling flexibility without adding parameters. 
Together, these modules preserve strong representational capacity under aggressive compression, enabling SSL-based speaker verification on mobile and embedded devices where attention layers are prohibitively expensive. 

SV-Mixer consistently demonstrates strong performance across various evaluation settings.
Compared to the Transformer encoder, it achieves a lower Equal Error Rate (EER) of 1.52\% on VoxCeleb1-O, which corresponds to a 14.6\% relative improvement. 
At the same time, it reduces the number of parameters by 55.4\% and the computational cost by 49.6\% per layer. 
It also outperforms alternative MLP-based architectures such as MLP-Mixer and CycleMLP by 16.5\% and 19.6\%, respectively, under identical training conditions. 
Under varying compression levels, SV-Mixer with all three proposed modules maintains robust performance and shows significantly better resilience to degradation compared to the vanilla model. 
Notably, at a 75\% compression ratio, it achieves performance nearly equivalent to the teacher model. 
In addition, SV-Mixer exhibits strong compatibility across different SSL teachers and speaker verification backends, further supporting its generality and deployment potential. 

The key contributions of this work are as follows:
\begin{itemize}
    \item \textbf{SV-Mixer.} We introduce \emph{SV-Mixer}, the first fully MLP-based student encoder for SSL distillation pipeline.
    Its three lightweight mixing modules jointly capture short-term and long-term temporal features and channel interactions within a compact footprint. 
    \item \textbf{Heterogeneous distillation.} This work is the first to show that a student architecture \emph{structurally dissimilar} to its Transformer teacher can be distilled effectively, challenging the prevailing assumption that student and teacher backbones must match for SSL distillation. 
    \item \textbf{On-device readiness.} By removing self-attention entirely, SV-Mixer reduces parameters and GMACs, lowering the computational barrier limiting SSL-based speaker verification on mobile and embedded platforms. 
\end{itemize}

\section{Related Work}
\subsection{SSL Model Compression}
Knowledge distillation has emerged as a primary approach for compressing large SSL models. 
Recent studies have proposed various distillation strategies tailored to speech models, aiming to reduce architectural complexity while preserving teacher-level representations. 
These methods typically distill information from a large Transformer-based model into a smaller student model under the same architectural paradigm. 
\textit{DistilHuBERT}~\cite{distilhubert} performs layer-wise distillation from HuBERT into a shallow student network with only two Transformer layers. 
It employs multiple prediction heads to match intermediate representations from different teacher layers. 
\textit{FitHuBERT}~\cite{fithubert} proposes a thin-and-deep student architecture that retains the teacher's depth while reducing its width. 
It further introduces hint-based distillation across all layers and temporal downsampling to improve inference speed. 
\textit{STaR}~\cite{star} relaxes the distillation constraint by matching temporal relations rather than feature values, using attention maps and Gram matrices to guide the student. 
While effective, these approaches maintain the Transformer architecture in the student model. 
As a result, they inherit the structural inefficiencies of self-attention, particularly its quadratic complexity, which limits inference scalability for long-form speech. 

Pruning-oriented approaches have also been explored.  
\textit{PARP}~\cite{parp} applies magnitude-based pruning to wav2vec 2.0 with iterative readjustment, whereas \textit{DPHuBERT}~\cite{dphubert} combines pruning and distillation to achieve better trade-offs between model size and performance.  
However, these methods often result in sparse and irregular structures that are not hardware-friendly and may degrade performance in complex tasks such as speaker verification. 

\subsection{Alternative Architectures to Transformers}
To move beyond the limitations of Transformer-based encoders, recent studies have explored alternative architectural paradigms. 
Among these, fully MLP-based models have gained attention for their simplicity, hardware efficiency, and potential scalability to longer sequences. 
Originally developed for image classification~\cite{mlpmixer}, the MLP-Mixer replaces self-attention with stacked multi-layer perceptrons (MLPs) that independently mix information along the token (spatial) and channel (feature) dimensions. 
Variants such as gMLP~\cite{gmlp} and ResMLP~\cite{resmlp} extend this architecture while retaining its attention-free nature. 
MLP-based models have also been adopted in speech applications such as keyword spotting and speaker verification~\cite{tang2024mlp, mlpsvnet}. 

In the context of SSL model compression, MLPs are attractive due to their linear computational complexity and hardware-friendly structure. 
However, in our preliminary experiments, standard architectures such as MLP-Mixer and CycleMLP consistently underperformed when directly used as student models in distillation pipelines. 
This observation suggests that these generic architectures are not well-suited to absorb teacher representations.

\section{Proposed Method}
We introduce \textbf{SV-Mixer}, a lightweight, attention-free student encoder tailored for SSL model distillation in speaker verification. 
Rather than modifying or scaling down existing Transformer-based students, SV-Mixer adopts a fundamentally different design based on multi-layer perceptrons, aiming to reduce both architectural complexity and deployment cost. 

While standard MLP architectures offer simplicity, they often struggle to replicate the representational power of Transformers—especially in distillation scenarios where fine-grained alignment with teacher features is critical. 
To address this, SV-Mixer integrates three speech-specialized mixing modules that compensate for the absence of self-attention by enhancing temporal resolution, contextual modeling, and spectral discrimination. 

The result is a student encoder that maintains high accuracy under compression while being more scalable than its Transformer-based counterparts. 
This section details the design of SV-Mixer, including its architectural components, mixing strategies, and knowledge distillation setup.

\begin{figure}
    \centering
    \includegraphics[width=1\columnwidth]{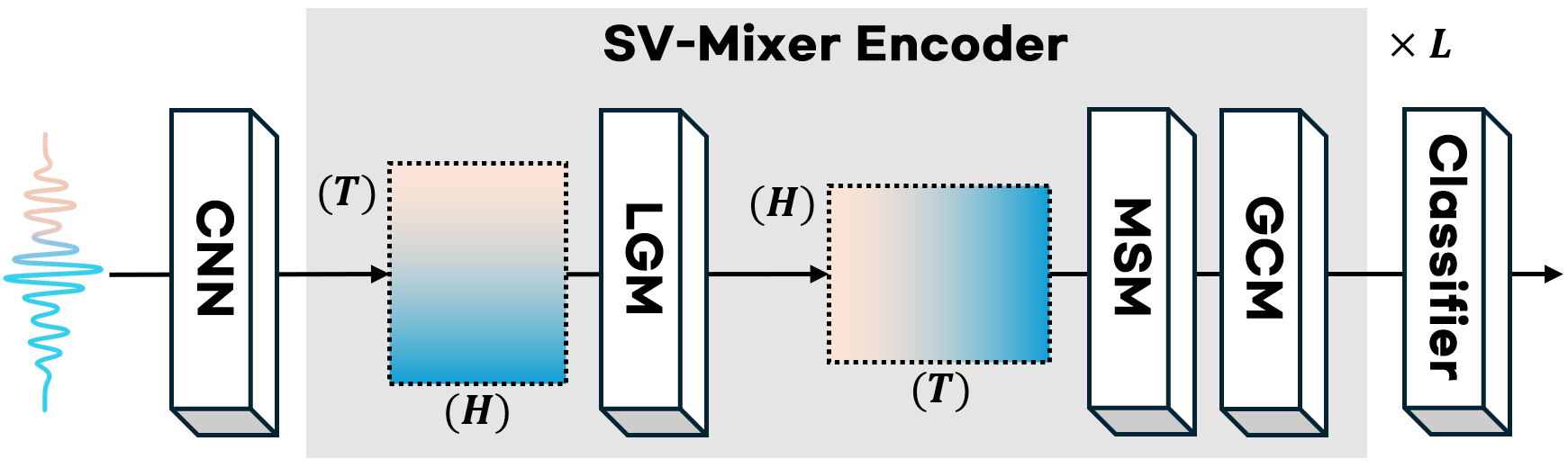}
    \vspace{-1em}
    \caption
    {
    Overall architecture of SV-Mixer. 
    The model consists of a convolutional frontend followed by stacked encoder blocks, each integrating Local-Global Mixing (LGM), Multi-Scale Mixing (MSM), and Group Channel Mixing (GCM) modules. 
    }
    \label{fig:overall}
    \vspace{-1em}
\end{figure}

\subsection{Overall Architecture}
SV-Mixer follows a modular architecture composed of a convolutional frontend, a stack of encoder blocks, and a classifier. 
The overall structure is illustrated in Figure~\ref{fig:overall}. 

The input waveform is first processed by a 7-layer 1D convolutional frontend that converts the raw audio into a frame-level feature sequence of shape $(T, H)$, where $T$ denotes the number of frames and $H$ the channel dimension. 
We adopt the same convolutional configuration used in the teacher model, with kernel sizes $\{10, 3, 3, 3, 3, 2, 2\}$, strides $\{5, 2, 2, 2, 2, 2, 2\}$, and 512 output channels per layer, to ensure a controlled comparison between Transformer and SV-Mixer. 

The resulting feature sequence is passed through $L$ encoder blocks. 
Whereas prior work has commonly adopted Transformer encoders at this stage, we instead employ SV-Mixer blocks designed to reduce architectural complexity and improve inference efficiency. 
Each SV-Mixer block consists of three stages: Local-Global Mixing, Multi-Scale Mixing, and Group Channel Mixing. 
Unlike Transformer encoders, SV-Mixer does not rely on positional encodings; instead, it implicitly learns temporal structure through its mixing mechanisms. 

To aggregate layer-wise information, we apply a learnable weighted sum over the outputs of all SV-Mixer blocks. 
The aggregated representation is then passed to a backend classifier, which in our case is a 512-dimensional ECAPA-TDNN~\cite{ecapa}, to produce speaker embeddings for verification. 
\begin{figure}
    \centering
    \includegraphics[width=1\columnwidth]{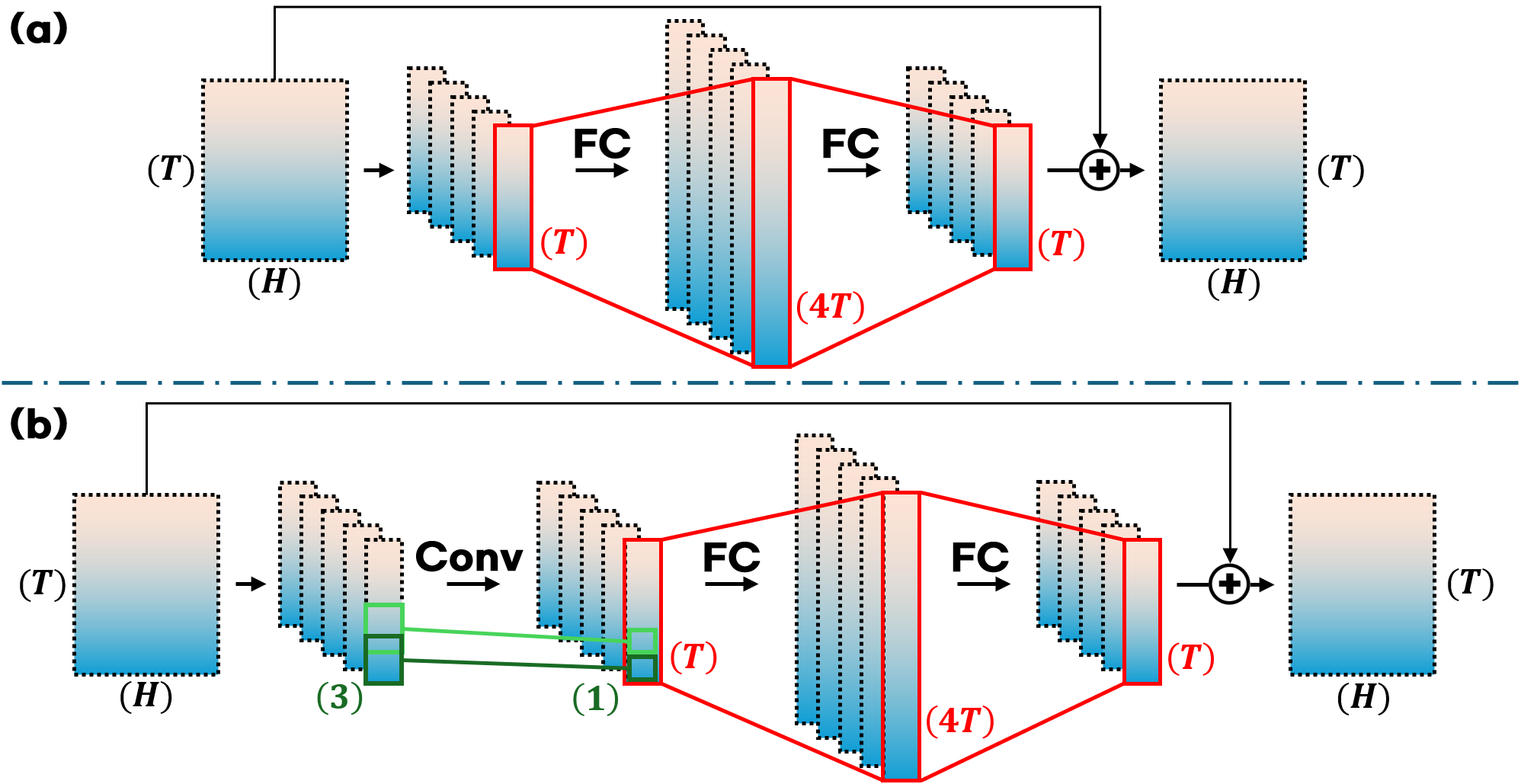}
    \vspace{-1em}
    \caption
    {
    (a) Mixing block from the original MLP-Mixer, composed of two fully connected (FC) layers for channel expansion and projection. 
    (b) Proposed LGM module that integrates temporal convolution and FC layers to capture both local and global features. 
    }
    \label{fig:normalmixing}
    \vspace{-1em}
\end{figure}

\subsection{SV-Mixer Encoder Architecture}
Each SV-Mixer block follows a modified MLP-Mixer design, adapted to better reflect the temporal and spectral characteristics of speech signals. 
The baseline MLP-Mixer performs two operations in sequence: token (temporal) mixing and channel mixing. 
As shown in Figure~\ref{fig:normalmixing} (a), each operation consists of two linear layers separated by a nonlinearity, with the first expanding the target dimension by a factor of four and the second projecting it back to the original size. 

Formally, a generic MLP layer is defined as:
\begin{equation}
\label{eq:normalmix}
y = \mathrm{GELU}(xW_1 + b_1)W_2 + b_2,
\end{equation}
where $x \in \mathbb{R}^T$ is the input, $W_1 \in \mathbb{R}^{T \times 4T}$ and $W_2 \in \mathbb{R}^{4T \times T}$ are weight matrices, and GELU is the activation function. 

To enhance the expressiveness of this structure for speech data, we replace the standard mixing blocks with three specialized modules: Local-Global Mixing (LGM), Multi-Scale Mixing (MSM), and Group Channel Mixing (GCM). 
These modules serve as drop-in replacements for the temporal and channel mixing layers in the original MLP-Mixer design. 

\begin{figure}
    \centering
    \includegraphics[width=1\columnwidth]{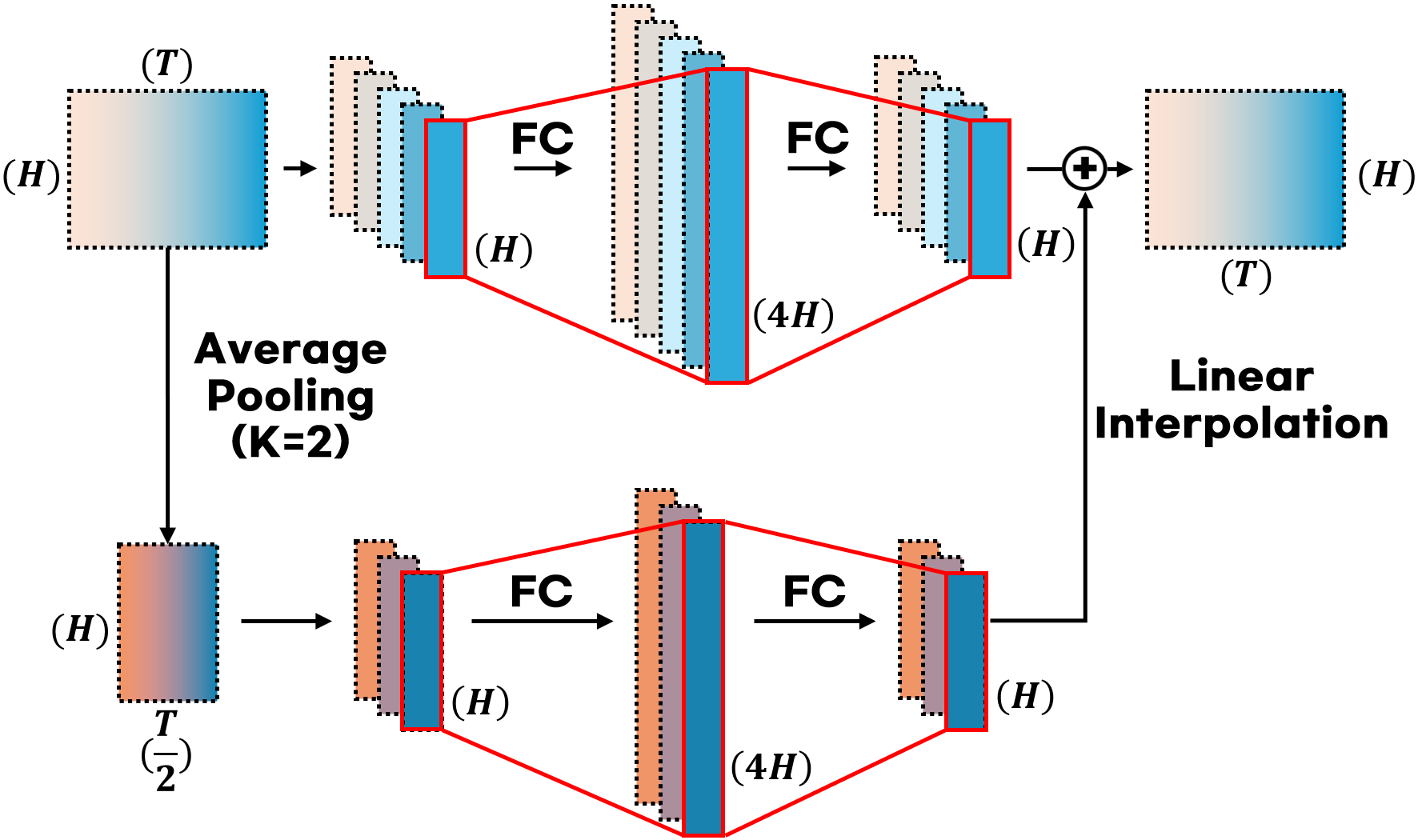}
    \vspace{-1em}
    \caption{
    Structure of the MSM module. 
    Parallel branches at original and low-resolution are merged to capture multi-scale temporal context. 
    }
    \label{fig:multiscale_mixing}
    \vspace{-1em}
\end{figure}

\begin{figure}
    \centering
    \includegraphics[width=1\columnwidth]{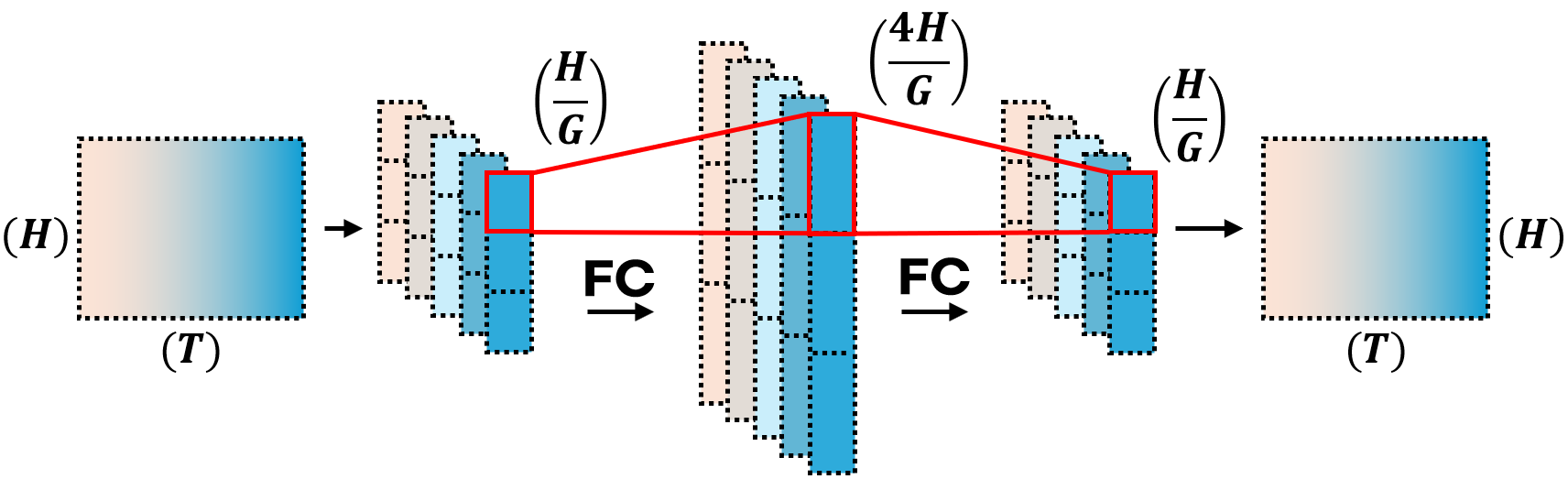}
    \vspace{-1em}
    \caption
    {
    Structure of the GCM module. 
    Grouped MLPs independently process channel subsets to capture localized interactions with fewer parameters.
    }
    \label{fig:groupfcmixing}
    \vspace{-1em}
\end{figure}

\paragraph*{\textbf{Local-Global Mixing}}
Figure~\ref{fig:normalmixing} (b) illustrates the overall structure of the LGM module. 
The Local-Global Mixing module enhances temporal modeling by combining the strengths of local convolutions and global representations. 
Instead of applying a single global MLP to all time steps, LGM adopts a two-stage strategy: a local 1D convolution first captures fine-grained dependencies among neighboring frames, followed by a global MLP that aggregates context across the full sequence. 
This separation allows the model to encode both short-term dynamics and long-range speaker characteristics. 
The design reflects common patterns in speaker verification systems, where local feature extraction and global pooling are used jointly to summarize utterance-level information~\cite{shim2022graph}. 
\paragraph*{\textbf{Multi-Scale Mixing}}  
Multi-Scale Mixing targets variability in temporal resolution by processing the input sequence at two scales simultaneously. 
As shown in Figure~\ref{fig:multiscale_mixing}, one branch operates at the original resolution, while the other applies average pooling $(\mathrm{kernel\ size} = 2)$ to obtain a downsampled view of the sequence. 
The low-resolution branch is then upsampled via linear interpolation and merged with the original-resolution output through element-wise addition. 
This dual-scale representation allows the model to encode speaker characteristics that manifest over different temporal spans, enhancing robustness to speaking rate and phonetic context variations. 
Previous studies have shown that such multi-resolution modeling improves speaker embedding quality~\cite{jung20d_interspeech}. 
\paragraph*{\textbf{Group Channel Mixing}}
Group Channel Mixing improves spectral modeling by introducing grouped channel-wise processing. 
Instead of applying a single MLP to the full channel dimension, GCM divides the input vector into $G$ disjoint groups and processes each group independently with a dedicated MLP, as illustrated in Figure~\ref{fig:groupfcmixing}. 
This grouped strategy enables localized transformations within spectral subspaces, allowing the model to focus on frequency-selective patterns that are known to be important for speaker recognition~\cite{ravanelli2018speaker}. 
It also draws on findings in multi-stream modeling for speech tasks, where group-wise processing improves both performance and parameter efficiency~\cite{yao2024multistream}. 
After independent processing, the group outputs are concatenated to form the final channel-mixed representation. 
In GCM, the input vector $x$ is split first into $G$ groups along the channel dimension. 
Each group is subsequently processed independently, as shown in Equation (2). 
\begin{align}
\label{eq:groupmix1}
y_g &= \mathrm{GELU}(x^{(g)}W_1^{(g)} + b_1^{(g)})W_2^{(g)} + b_2^{(g)}, \nonumber \\ 
&\quad g \in \{1, 2, ..., G\},
\end{align}
where $x \in \mathbb{R}^{H/G}$ is the input for group $g$. The group outputs are then concatenated to form the final output:
\begin{equation}
\label{eq:groupmix2}
y = \mathrm{Concat}(y_1, y_2, ..., y_G).
\end{equation}

\subsection{Knowledge Distillation Strategy}
SV-Mixer is trained under the OS-KDFT framework~\cite{sv2}, which performs knowledge distillation and task-specific fine-tuning simultaneously. 
In this setup, the student model is optimized to match the intermediate representations of a frozen teacher model while also learning to discriminate speakers via a supervised loss. 

A key feature of OS-KDFT is its forward-only distillation strategy: the teacher network is kept in inference mode throughout training, and no gradients are propagated through it. 
This significantly reduces memory and computational overhead, making the framework well suited for small-scale research hardware or mobile development environments. 

For distillation, we apply Mean Squared Error (MSE) loss between the hidden state outputs of the student and teacher layers. 
In addition to the distillation loss, we use the Additive Angular Margin (AAM) Softmax~\cite{aam} loss to directly optimize speaker separability. 
To further improve learning from hard examples, we incorporate a top-$K$ hard sample penalty within each batch, where the five most confusing impostor samples are penalized with a weight multiplier of 10. 

This dual-objective formulation allows SV-Mixer to benefit from the representational strength of the teacher model while adapting to the speaker verification task. 
By avoiding gradient flow through the teacher, the method maintains low training complexity and improves practical applicability in resource-constrained scenarios. 

\begin{table*}[th!]
\caption{
Comparison of encoder architectures in terms of EER (\%) and MinDCF on both generalization (Vox1-O/E/H, VoxSRC23) and environmental mismatch (VCMix, VOiCES) test sets. 
Size and GMACs are reported per layer.
}
\label{table:compare}
\centering
\resizebox{\textwidth}{!}{
\begin{tabular}{ccc|cccccccc|cccc} 
\toprule
\multirow{3}{*}{Encoder} & \multirow{3}{*}{\makecell{Size\\/Layer}} & \multirow{3}{*}{\makecell{GMACs\\/Layer}} & \multicolumn{8}{c|}{Generalization Set} & \multicolumn{4}{c}{Environmental Mismatch Set} \\
\addlinespace[0.5ex]
\cline{4-15}
\addlinespace[0.5ex]
 & & & \multicolumn{2}{c}{Vox1-O} & \multicolumn{2}{c}{Vox1-E} & \multicolumn{2}{c}{Vox1-H} &
\multicolumn{2}{c|}{VoxSRC23} & \multicolumn{2}{c}{VCMix} & \multicolumn{2}{c}{VOiCES} \\
 & & & EER & MinDCF & EER & MinDCF & EER & MinDCF & EER & MinDCF & EER & MinDCF & EER & MinDCF \\
\midrule
Transformer & 8.40M & 1.25 & 1.78 & 0.123 & 1.92 & 0.122 & 3.47 & 0.212 & 7.80 & 0.415 & 6.25 & 0.385 & 12.03 & 0.486 \\
MLP-Mixer & 8.58M & 1.43 & 1.82 & 0.125 & 1.78 & 0.112 & 3.13 & 0.192 & 6.77 & 0.366 & \textbf{4.84} & 0.343 & \textbf{11.13} & \textbf{0.480} \\
CycleMLP & 9.45M & 1.42 & 1.89 & 0.128 & 1.86 & 0.117 & 3.28 & 0.199 & 6.91 & 0.395 & 5.30 & 0.388 & 12.77 & 0.529 \\
ConvMixer & 9.02M & 1.48 & 1.76 & 0.122 & 1.75 & 0.120 & 3.05 & 0.189 & 6.68 & 0.372 & 5.32 & 0.414 & 11.91 & 0.505 \\
SV-Mixer & \textbf{3.75M} & \textbf{0.63} & \textbf{1.52} & \textbf{0.110} & \textbf{1.64} & \textbf{0.104} & \textbf{2.98} & \textbf{0.186} & \textbf{6.61} & \textbf{0.362} & 4.96 & \textbf{0.321} & 12.24 & \textbf{0.480} \\
\bottomrule
\end{tabular}
}
\vspace{-1em}
\end{table*}

\section{Experiments}
\subsection{Dataset}
All student models are trained on the VoxCeleb2 development set~\cite{vox2}, a large-scale corpus commonly used for speaker verification. 
To improve robustness and generalization, we apply data augmentation using MUSAN~\cite{musan} for additive noise and music, as well as simulated room impulse responses for reverberation~\cite{rir}.

Evaluation is performed on the VoxCeleb1~\cite{vox1} test sets, including VoxCeleb1-O (original), VoxCeleb1-E (extended), and VoxCeleb1-H (hard), following the standard protocol.  
To further assess cross-domain generalization, we evaluate on the VoxSRC 2023 validation set~\cite{voxsrc23} and two domain-mismatched datasets: VCMix~\cite{vcmix} and VOiCES~\cite{voices}. 
\begin{table}[t]
\caption{
Ablation study on the impact of the three proposed mixing modules. 
Token or channel mixing in a vanilla MLP-Mixer is replaced by LGM, GCM, or MSM. 
}
\label{table:ablation}
\centering
\resizebox{0.95\linewidth}{!}{
\begin{tabular}{ccc|cccc} 
\toprule
GCM & LGM & MSM & Vox1-O & Vox1-E & Vox1-H & VoxSRC23 \\
\midrule
& & & 1.82 & 1.78 & 3.13 & 6.77 \\
\midrule
\checkmark & & & 1.62 & 1.69 & 3.10 & \textbf{6.59} \\
& \checkmark & & 1.64 & 1.72 & 3.18 & 6.90 \\
& & \checkmark & 1.75 & 1.74 & 3.25 & 6.73 \\
\midrule
\checkmark & \checkmark & \checkmark & \textbf{1.52} &\textbf{ 1.64} & \textbf{2.98} & 6.61 \\
\bottomrule
\end{tabular}
}
\vspace{-1em}
\end{table}

\subsection{Baseline}
To evaluate the effectiveness of SV-Mixer as a student encoder, we compare it against a Transformer-based baseline under controlled conditions. 
Both models are trained using the same WavLM-Large teacher, convolutional frontend, and backend classifier. 
The only difference lies in the encoder architecture. 
All other components, including the hidden dimensionality and training objectives, are held constant to isolate the impact of encoder architecture.

\subsection{Implementation Details}
All models are optimized using the AdamW optimizer~\cite{adamw} with a weight decay of $2 \times 10^{-5}$. 
The initial learning rate is set to $2 \times 10^{-4}$ and is halved if validation performance does not improve for five consecutive epochs.  
Early stopping is triggered if no improvement is observed for 10 epochs.

During training, input waveforms are randomly cropped to 3 seconds, and the batch size is set to 128. 
All experiments were performed using two NVIDIA RTX A6000 GPUs with PyTorch. 
To promote reproducibility, we make our code, pretrained models, and inference scripts publicly available on GitHub \footnote{https://github.com/Jungwoo4021/SV-Mixer}. 

\section{Results}
\subsection{Comparison of Encoder Architectures}
We compare the performance of various student encoder architectures distilled from a common WavLM-Large teacher under identical training settings. 
As shown in Table~\ref{table:compare}, the proposed SV-Mixer consistently outperforms the baseline Transformer encoder across all standard VoxCeleb1 evaluation sets. 
On VoxCeleb1-O, SV-Mixer reduces the Equal Error Rate (EER) from 1.78\% to 1.52\%, corresponding to a 14.6\% relative improvement. 
Similar gains are observed on VoxCeleb1-E and VoxCeleb1-H, where SV-Mixer achieves relative EER reductions of 14.6\% and 14.1\%, respectively. 
On the VoxSRC 2023 validation set, SV-Mixer yields an EER of 6.61\%, compared to 7.80\% for the Transformer baseline. 
SV-Mixer also shows a 20.6\% relative reduction on the VCMix dataset (4.96\% vs. 6.25\%), although a slight degradation is observed on VOiCES. 

In terms of model efficiency, SV-Mixer is substantially more lightweight. 
Each SV-Mixer block contains 55.4\% fewer parameters and requires 49.6\% fewer GMACs per layer than a Transformer block. 

We also evaluate other MLP-based architectures such as MLP-Mixer~\cite{mlpmixer}, CycleMLP~\cite{cyclemlp}, and ConvMixer~\cite{convmixer}. 
While these models perform comparably to the Transformer baseline when trained from scratch, they underperform in the distillation setting. 
This suggests that generic MLP architectures are less effective at absorbing knowledge from a Transformer teacher, and highlights the importance of task- and domain-specific adaptation in student design. 

\begin{figure}
    \centering
    \includegraphics[width=1\columnwidth]{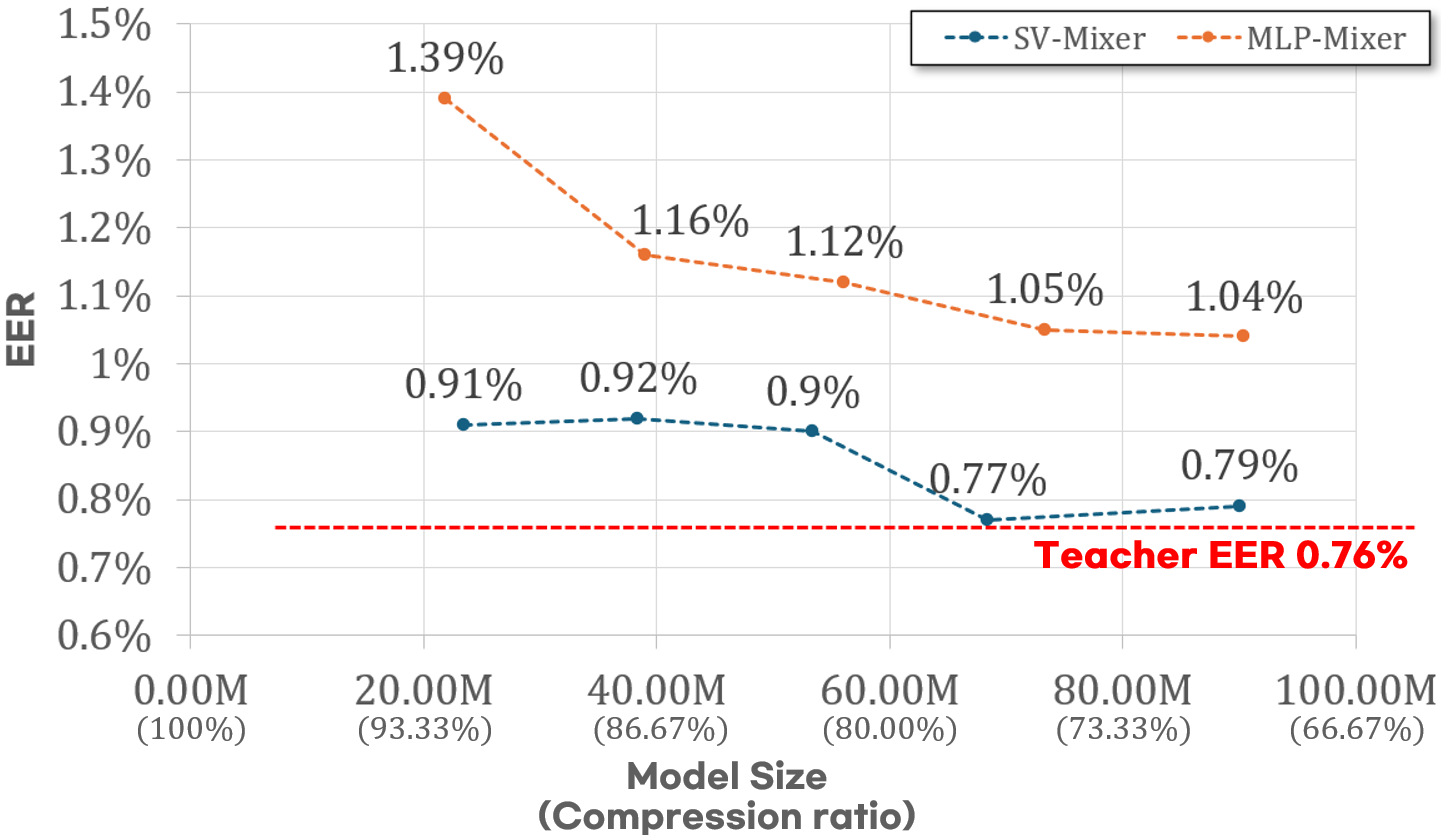}
    \vspace{-1em}
    \caption{
    Compression–performance trade-off for SV-Mixer and MLP-Mixer.
    Each point represents a student model with a specific size and compression ratio relative to the WavLM-Large teacher. EER is measured on VoxCeleb1-O. 
    Distillation follows the DistilHuBERT strategy.
    }
    \label{fig:compression}
    \vspace{-1em}
\end{figure}

\subsection{Ablation Study}
To better understand how each mixing module contributes to performance, we conduct an ablation study by selectively employing GCM, LGM, and MSM from the SV-Mixer architecture. 
Table~\ref{table:ablation} summarizes the results on VoxCeleb and VoxSRC23 test sets. 
We begin with a baseline variant in which all three modules are removed and replaced with the standard token and channel mixing blocks from MLP-Mixer. 
This configuration achieves an EER of 1.82\% on VoxCeleb1-O and 6.77\% on VoxSRC23, establishing a lower bound for performance. 
Among the individual modules, GCM provides the most consistent improvement across datasets, reducing EER to 1.62\% on VoxCeleb1-O and to 6.59\% on VoxSRC23. 
LGM contributes significantly on VoxCeleb1-O (1.67\%), but degrades performance on the cross-domain set. 
Although MSM alone yields limited improvements, it contributes meaningfully when combined with the other modules. 
When all three modules are included, SV-Mixer achieves the best overall performance: 1.52\% on VoxCeleb1-O and 6.61\% on VoxSRC23. 
These results confirm that the three proposed mixing strategies are complementary and contribute jointly to SV-Mixer’s ability to generalize across both matched and mismatched domains. 
\begin{table}[t]
\caption{
EER (\%) for different encoder–backend combinations on the generalization test sets.
}
\label{table:backend}
\centering
\resizebox{\linewidth}{!}{
\begin{tabular}{c|c|cccc} 
\toprule
Classifier & Encoder type & Vox-O & Vox-E & Vox-H & VoxSRC23 \\
\midrule
\multirow{2}{*}{Linear} & Transformer & 2.48 & 2.28 & 3.70 & 8.13 \\
& SVMixer & \textbf{2.00} & \textbf{1.96} & \textbf{3.25} & \textbf{7.59} \\
\midrule
\multirow{2}{*}{ECAPA} & Transformer & 1.78 & 1.92 & 3.47 & 7.80 \\
& SVMixer & \textbf{1.52} & \textbf{1.64} & \textbf{2.98} & \textbf{6.61} \\
\midrule
\multirow{2}{*}{RedimNet} & Transformer & 1.66 & 1.63 & 2.74 & 6.71 \\
& SVMixer & \textbf{1.51} & \textbf{1.47} & \textbf{2.54} & \textbf{6.08} \\
\bottomrule
\end{tabular}
}
\vspace{-1em}
\end{table}

\subsection{Compression Robustness}
We conduct experiments using student models with different parameter budgets to evaluate the robustness of SV-Mixer under varying compression levels. 
In contrast to our default MSE loss on the final hidden states, this analysis leverages the DistilHuBERT method~\cite{distilhubert}, which supports deeper compression via intermediate-layer supervision. 
To isolate the effectiveness of our proposed design, this experiment compares SV-Mixer to MLP-Mixer rather than the Transformer-based model. 
Figure~\ref{fig:compression} presents the Equal Error Rate on VoxCeleb1-O plotted against student model size, expressed in millions of parameters. 
Compression ratios are reported relative to the WavLM-Large teacher model. 
Across all model sizes, SV-Mixer consistently outperforms the baseline MLP-Mixer, indicating superior compatibility with Transformer-based knowledge distillation.  
Notably, SV-Mixer maintains strong performance even at high compression levels: at a 70–80\% reduction in parameter count, it achieves an EER of 0.77\%, nearly matching the teacher model's performance of 0.76\%. 
In contrast, the MLP-Mixer exhibits a steep degradation in accuracy as model size decreases.
These results demonstrate that SV-Mixer not only improves performance in the standard distillation setting, but also exhibits greater resilience to architectural downsizing, making it a promising candidate for deployment in resource-constrained environments.

\subsection{Effect of Backend Classifier}
To assess the compatibility of SV-Mixer with various speaker verification pipelines, we evaluate its performance under different backend classifiers of increasing complexity.  
We compare three backends: a single linear layer, the ECAPA-TDNN~\cite{ecapa}, and ReDimNet~\cite{redimnet}, all trained using AAM-Softmax loss.
Table~\ref{table:backend} reports the Equal Error Rates (EERs) of both SV-Mixer and the Transformer baseline across VoxCeleb and VoxSRC23. 
SV-Mixer consistently outperforms the Transformer-based student across all backends. 
Notably, the performance gap is more pronounced with lightweight classifiers. 
For instance, when paired with a linear layer, SV-Mixer achieves an EER of 2.00\% on VoxCeleb1-O and 7.59\% on VoxSRC23, compared to 2.48\% and 8.13\% with the Transformer. 
These results suggest that SV-Mixer produces more discriminative embeddings even in the absence of complex backend modeling, which is especially advantageous in resource-limited scenarios.  
Performance improves consistently with more expressive backends such as ECAPA-TDNN and ReDimNet, suggesting that SV-Mixer scales well with backend capacity without observable saturation. 

\begin{table}[t]
\caption{
Performance of Transformer and SV-Mixer distilled from various SSL teacher models. 
}
\label{table:teacher}
\centering
\resizebox{\linewidth}{!}{
\begin{tabular}{c|c|cccc} 
\toprule
Classifier & Encoder type & Vox-O & Vox-E & Vox-H & VoxSRC23 \\
\midrule
\multirow{2}{*}{Wav2Vec2.0} & Transformer & 3.18 & 3.34 & 5.72 & 11.24 \\
& SVMixer & \textbf{2.23} & \textbf{2.21} & \textbf{3.65} & \textbf{7.83} \\
\midrule
\multirow{2}{*}{HuBERT} & Transformer & 1.71 & 1.70 & 3.21 & 6.88 \\
& SVMixer & \textbf{1.63} & \textbf{1.69} & \textbf{3.08} & \textbf{6.74} \\
\midrule
\multirow{2}{*}{WavLM-Large} & Transformer & 1.78 & 1.92 & 3.47 & 7.80 \\
& SVMixer & \textbf{1.52} & \textbf{1.64} & \textbf{2.98} & \textbf{6.61} \\
\midrule
\multirow{2}{*}{WavLM-Base+} & Transformer & 1.89 & 1.88 & 3.76 & 7.92 \\
& SVMixer & \textbf{1.63} & \textbf{1.63} & \textbf{2.98} & \textbf{6.52} \\
\bottomrule
\end{tabular}
}
\vspace{-1em}
\end{table}

\subsection{Effect of Teacher Model}
We further evaluate the generalizability of SV-Mixer by pairing it with different teachers during distillation. 
Specifically, we conduct experiments using four teacher models: WavLM Base+, WavLM Large, HuBERT, and wav2vec 2.0. 
Table~\ref{table:teacher} summarizes the performance of SV-Mixer and the Transformer-based student across the VoxCeleb and VoxSRC23 evaluation sets. 
While absolute performance varies depending on the teacher’s capacity and pretraining data, SV-Mixer consistently achieves lower EERs than the Transformer student across all teacher variants. 
For instance, when distilled from wav2vec 2.0, SV-Mixer reduces the EER on VoxCeleb1-O from 3.18\% to 2.23\%. 
Similarly, with WavLM Base+, SV-Mixer achieves 1.63\% on VoxCeleb1-O and 6.52\% on VoxSRC23, outperforming the baseline by notable margins. 
These findings indicate that SV-Mixer is robust to differences in teacher pretraining objectives and size. 
Its effectiveness is not tied to a specific teacher, making it broadly applicable across diverse self-supervised learning models. 

\section{Conclusion}
This paper presents SV-Mixer, a lightweight, attention-free encoder for self-supervised speaker verification via knowledge distillation. 
Replacing Transformer layers with speech-oriented MLP mixing modules, SV-Mixer delivers strong performance under aggressive compression. 
Extensive experiments show that it outperforms Transformer and other MLP students across in-domain and cross-domain tests while cutting model size and compute. 
SV-Mixer remains robust to teacher models, backend classifiers, and compression levels, making it suitable for resource-constrained deployment. 
Our study has several limitations. 
In particular, it adopts a fixed training setup and employs only a single distillation strategy, potentially overlooking more effective combinations of training objectives and transfer mechanisms. 
Future work could explore broader combinations of student architectures~\cite{redimnet}, loss functions~\cite{wensphereface2}, and knowledge distillation methods~\cite{star, jin24c_interspeech} to further advance the state of the art in self-supervised speaker verification.

\bibliographystyle{IEEEtran}
\bibliography{refs}

\end{document}